\documentclass{article}

\usepackage{todonotes}
\usepackage{xspace}
\usepackage{amsmath}
\newcommand{\gnet}{\textit{GermlineNet}\xspace}
\newcommand{\snet}{\textit{SomaticNet}\xspace}
\usepackage{float}
\usepackage{subfig}

\PassOptionsToPackage{numbers, compress}{natbib}
\usepackage[final]{nips_2018}

\usepackage[utf8]{inputenc} % allow utf-8 input
\usepackage[T1]{fontenc}    % use 8-bit T1 fonts
\usepackage{hyperref}       % hyperlinks
\usepackage{url}            % simple URL typesetting
\usepackage{booktabs}       % professional-quality tables
\usepackage{amsfonts}       % blackboard math symbols
\usepackage{nicefrac}       % compact symbols for 1/2, etc.
\usepackage{microtype}      % microtypography

\author{
  Luke R Harries$^1$, Suyi Zhang$^1$, Geoffroy Dubourg-Felonneau$^1$,  James H R Farmery$^1$, \\
  \textbf{Jonathan Sinai$^1$, Belle Taylor$^1$, Nirmesh Patel$^1$, John W Cassidy$^1$$^,$$^2$, John Shawe-Taylor$^3$,} \\
  \textbf{Harry W Clifford$^1$} \\
  \\
  \and
  $^1$Cambridge Cancer Genomics \\
  $^2$University of Cambridge \\
  $^3$University College London
}

\title{Interlacing Personal and Reference Genomes for Machine Learning Disease-Variant Detection}

\begin{document}

\maketitle

\begin{abstract}
DNA sequencing to identify genetic variants is becoming increasingly valuable in clinical settings. Assessment of variants in such sequencing data is commonly implemented through Bayesian heuristic algorithms. Machine learning has shown great promise in improving on these variant calls, but the input for these is still a standardized "pile-up" image, which is not always best suited. In this paper, we present a novel method for generating images from DNA sequencing data, which interlaces the human reference genome with personalized sequencing output, to maximize usage of sequencing reads and improve machine learning algorithm performance. We demonstrate the success of this in improving standard germline variant calling. We also furthered this approach to include somatic variant calling across tumor/normal data with Siamese networks. These approaches can be used in machine learning applications on sequencing data with the hope of improving clinical outcomes, and are freely available for noncommercial use at \url{www.ccg.ai}.
\end{abstract}

\section{Introduction}

Genetic aberrations are defects in DNA which have been linked, in many cases, to human ill-health. Heritable (germline) variants have been shown to be directly responsible for many genetic disorders and disease predispositions worldwide, in both simpler monogenic inheritance disorders such as cystic fibrosis \cite{Ratjen2003CysticF}, sickle cell anaemia \cite{Higgs2008GeneticCI} and thalassaemia \cite{Chen1997MolecularDO}, and in complex polygenic inheritance disorders such as Alzheimer’s \cite{Cauwenberghe2016TheGL} and autoimmune diseases \cite{Ramos2015GeneticsOA}. Additionally, mutations in a humans post-conception (i.e. a somatic variant) can also cause disease. Cancer is caused by such somatic mutations and is common worldwide, accounting for an estimated 8.2 million deaths in 2012 \cite{doi:10.1002/ijc.29210}.

In clinical settings, massively parallel personalized DNA sequencing is becoming commonplace, with Genomics England, for example, pledging to sequence five million Britons over the next five years \cite{GenomicsEngland}. Such efforts will enable broad coverage of the genome for identification of rare variants. Similarly, the sequencing of cancers is gaining popularity due to their genetic complexity and heterogeneity, with any given tumor being genetically unique. This personalized sequencing information can improve prevention, diagnosis, and treatment with precision medicine approaches \cite{Mardis2009CancerGS}.

Genome sequencing is a rich source of data, and a clear target for machine learning approaches and algorithms. Traditionally, assessment of variants and mutations in DNA sequencing data is implemented through Bayesian heuristic algorithms \cite{Liu2013VariantCF}. More recently, machine learning approaches have started to demonstrate improved results. For example, DeepVariant, an analysis pipeline developed by Google, recently won the PrecisionFDA Truth Challenge through neural network detection of germline variants \cite{Precisio50:online,Poplin2018CreatingNetworks}. These approaches rely on a generated image of the sequencing reads known as a pile-up image. Pile-ups are commonly produced in bioinformatics research to visualize read alignments at specific genomic loci, for example in the Integrative Genomics Viewer \cite{Robinson2011IntegrativeGV}. However, this may not be best suited to machine learning applications. Population-scale data, like the human reference genome, is important in understanding how the individual's personalized sequence-output differs from the general population. This reference genome is not well emphasized in the traditional pile-up and is confined to a small portion of the image.

In this paper, we describe our efforts to adapt DNA sequencing pile-up generation to build an improved input for machine learning algorithms. We do this by interlacing reference genome sequence with personalized sequencing output to produce a spliced image. Firstly, we demonstrate the value of this through our own germline variant caller (GermlineNET), for which the spliced kernal exhibits improved results. Secondly, we incorporate this into a new architecture for somatic variant calling (SomaticNET), which utilizes a novel method to analyze across tumor/normal paired sequencing data with Siamese networks. These tools are freely available for noncommercial use at \url{www.ccg.ai}.

\section{Methods}

\subsection{Data Generation}

For germline variant data we used one of the "Platinum genomes": a series of well studied genomes originally sequenced as part of the 1,000 genomes project \cite{Schafer2010AMO}. This included serial sequencing of immediate family members to enable haplotype reconstruction and high confidence variant identification \cite{Eberle2017ARD}. The 17 member CEPH pedigree 1463 was sequenced to 50x depth on a HiSeq 2000 system, of which we utilized the well characterized NA12878 sample. For somatic variant detection, we utilized a dataset generated through BAMsurgeon \cite{Ewing2015CombiningTG}, a tool designed to "spike-in" artificial mutations in a sequencing file. Somatic variant callers run on a tumor/normal pair of files; for our data set we utilized a pair of independently sequenced samples from the same individual, with BAMsurgeon applied to one sample output, to add known variants and form a synthetic tumor.

\subsection{Data Pre-processing}

GermlineNET and SomaticNET shared a data pre-processing pipeline similar to that of Google's DeepVariant \cite{Precisio50:online,Poplin2018CreatingNetworks}, which generates pile-up images from the raw DNA sequencing data. This pile-up image is then RGB-encoded and used as input to the CNNs for classification. This process was implemented over two modules. The first module identifies variant candidate sites using a frequency heuristic. The human genome contains 3 billion base pairs, in which germline variants occur at approximately one in every 300 bases \cite{Sham2011GeneticDiseases}. In order to reduce the number of sites the CNN needs to classify, we use an initial filter with high sensitivity and low specificity. We used the same simple heuristic---Variant Allele Frequency (VAF, percentage of mismatched bases given a location) threshold of 5\%, following DeepVariant's approach. The second module creates RGB-encoded pile-up images for each candidate site. Base sequences from reads overlapping candidate sites identified earlier are encoded as tensor objects in the first channel. A second channel is used to encode base quality scores. In theory, additional channels can be added to represent information from the sequencing process, such as DNA strand (positive or negative) and mapping quality scores (MAPQs).

\subsection{Spliced Pile-up}
The purpose of the pile-up image is to allow a direct comparison of DNA fragments from the dataset with the reference, to identify variants. Traditionally, a pile-up displays a section of the human reference genome along the top of the image, together with the sequenced fragments of DNA from the given dataset that align to this region (explained further in Appendix A.1). However, as the CNN can only extract features which the kernel is applied to, the reference DNA is then only compared with the normal DNA at the border. We introduce a novel method for improving pile-up image input and maximizing feature-use through a design called the \textit{spliced pile-up}. The suggested \textit{spliced pile-up} approach combines the reference DNA and the sequenced DNA for the first CNN layer, keeping the reference at a constant position in the kernel. This allows the Kernel to learn the weights needed to appropriately combine the two pieces of information (reference and sequenced DNA). We achieve this by altering input images such as to insert the reference at every $s$ rows, where $s$ is the vertical stride of the convolutional kernel ($s > 1$). An example of these \textit{spliced images} is shown in Figure 1.1.

\begin{figure}[H]
\centering
\subfloat{\includegraphics[width=6.9cm]{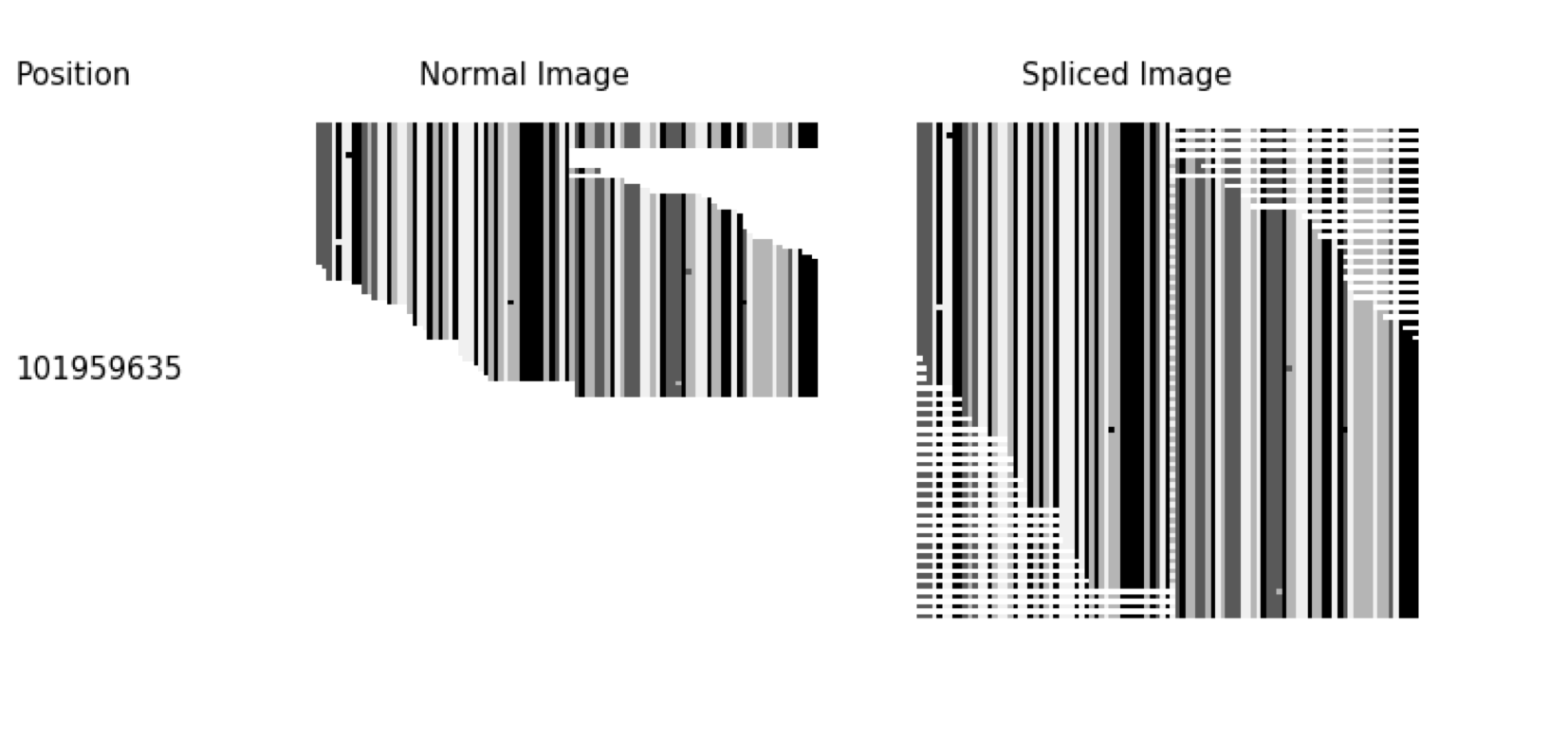}}
\subfloat{\includegraphics[width=5.5cm]{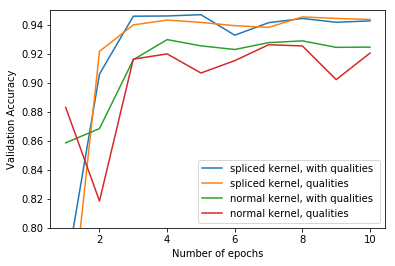}}
\caption{(1) Example Normal and Spliced images, with the latter containing the reference at alternating lines. (2) Number of epochs vs validation loss of ResNet34 for the first fold of cross-validation. The spliced image learns quicker and reaches higher accuracy.}
\end{figure}

\subsection{CNN Architecture and Training}
GermlineNET uses a CNN to classify the difference between the sequenced DNA and the reference sequence in pile-up images as being (a) an error in sequencing or (b) a germline variant. Inception and ResNET were candidate state-of-the-art network architectures for image classifications \citep{Szegedy2015RethinkingVision,He2015DeepRecognition}. We used the softmax as a final layer to calculate class probabilities, and the negative log-likelihood as the loss function for gradient descent. For model training, we used a balanced dataset of 48,000 true positive pile-up images from ground truth, and 48,000 images from other random locations. Training and validation sets were split across chromosomes (training: 1-17, testing: 18-22 and X).

For somatic variant calling, our novel approach utilizes a Siamese CNN \cite{ChopraLearningVerification}. SomaticNET takes two images as inputs, feeds them forward through the neural network to give two vectors in latent space, and returns a function applied to the distance between the two vectors. We use the euclidean distance between the vectors and train with the Contrastive Loss Function, where distance smaller than a threshold value indicates (a) no mutation present in the tumor candidate site, while a larger distance indicates (b) a mutation present in the tumor candidate site. The weights from GermlineNET are used to initialize those in SomaticNET, making the two sister networks. We used a balanced dataset of 22,875 true positive pile-up images, and 22,875 images from other random locations.

% \snet's neural network is a Siamese convolutional neural network. The Siamese neural network (introduced in Section \ref{sec:siamese}) takes two vectors as inputs, feeds them forward through the same neural network giving two vectors in latent space, and returns a function applied to the distance between the two vectors. In \snet's Siamese CNN, the sister network is \gnet's CNN, with the weight initialised from SNP training.

% The two sister networks each output a vector of size $4 \times 1$. In order to use the Siamese network as a classifier, we use the euclidean distance between the two vectors and trained with the Contrastive Loss Function (Section \ref{sec:siamese}), to learn that the two vectors should have no distance between them when there is (0) no mutation present in the tumor sample's candidate site, and that there should be a large distance between them when there is (1) a mutation present in the tumor sample's candidate site. 

% Additionally, we tested using calculated the pairwise distance between the two vectors followed by a fully connected layer and the sigmoid function. This resulted in a scalar output. An output of 0 was defined to mean they were of the same type---the differences at the candidate sites were errors. An output of 1 was defined to mean they were different---the differences at the candidate sites were due to a somatic variant. However, this resulted in a much lower training and validation accuracy.

\section{Results}

\subsection{GermlineNET}

Different model architectures and the addition of the spliced pile-up were tested with GermlineNET using K-fold cross-validation (K=4). Test statistics show the average of the best model of each cross-validation cycle, using F1 score (Table \ref{table:crossfoldsnp}). 20 epochs of training was conducted in each cycle and statistics were calculated on the test fold. The architecture ResNet34 resulted in the highest F1 (93.7\%), and using the spliced pile-up and quality improved this further to 95.3\% (Figure 1.2). Interestingly, performance in the architecture InceptionV4 was higher without the spliced pile-up, which may be due to the complex nature of this network resulting in ineffective backpropagation to the first layer, and should be further assessed on similar inception models.

\begin{table}[h]
\centering
\begin{tabular}{lllccccc}
\toprule
Model & Spliced & Quality &  Accuracy &     F1 &    ROC AUC &  Recall &  Precision \\
\midrule
\textbf{ResNet34} & False & False &      93.7 &  93.7 &  98.2 &    94.2 &       93.3 \\
         &       & True  &      94.0 &  94.1 &  98.6 &    94.8 &       93.6 \\
         & \textbf{True}  & False &      \textbf{95.3} &  95.3 &  98.9 &    \textbf{95.7} &       94.9 \\
         &       & \textbf{True}  &      \textbf{95.3} &  \textbf{95.4} &  \textbf{99.0} &    95.6 &       \textbf{95.2} \\
Inception v4 & False & False &      76.8 &  81.8 &  82.0 &    93.7 &       74.7 \\
         &       & True  &      85.7 &  85.7 &  92.7 &    86.0 &       85.6 \\
         & True  & False &      67.9 &  71.8 &  74.6 &    79.2 &       67.7 \\
         &       & True  &      81.6 &  81.4 &  88.8 &    81.3 &       81.9 \\
\bottomrule
\end{tabular}
\caption{Results of K-fold cross-validation for \gnet's neural network (top results in bold)}\label{table:crossfoldsnp}
\end{table}

\subsection{SomaticNET}

We evaluated SomaticNET's performance using similar cross-validation procedures, with the effects of model architectures and the spliced pile-up summarized in Table \ref{table:crossfoldsnv}. The Siamese network using ResNet34 showed better performance with spliced pile-up (F1 score 93.6\%). In this case, including quality scores led to deterioration in model performance, most likely because the semi-simulated data had synthetic mutated variants that do not correlate with data quality scores.

Distribution of Siamese distance on validation data showed SomaticNET can distinguish between tumor and normal tissue classes (Figure 2.1). The Siamese distance used for variant calling can be optimized through a grid search, with high validation accuracy, sensitivity and precision (Figure 2.2).

\begin{table}[h]
\centering
\begin{tabular}{lllccccc}
\toprule
Model & Spliced & Quality &  Accuracy &        F1 &       ROC AUC &    Recall &  Precision \\
\midrule
\textbf{ResNet34} & False & False &      92.5 &  92.2 &  96.2 &    89.6 &       95.4 \\
         &       & True  &      91.5 &  90.9 &  96.8 &    86.6 &       96.2 \\
         & \textbf{True}  & \textbf{False} &      \textbf{93.7} &  \textbf{93.6} &  95.9 &    \textbf{93.1} &       94.5 \\
         &       & True  &      92.1 &  91.6 &  \textbf{97.1} &    87.8 &       \textbf{96.3} \\
Inception v4 & False & False &      51.6 &  44.0 &  52.2 &    40.7 &       52.7 \\
         &       & True  &      51.9 &  27.8 &  52.2 &    20.1 &       55.2 \\
         & True  & False &      51.8 &  32.9 &  53.2 &    26.5 &       55.3 \\
         &       & True  &      51.4 &  26.8 &  53.2 &    18.8 &       54.0 \\
\bottomrule
\end{tabular}
\caption{Results of K-fold cross-validation for \snet's neural network (top results in bold)}\label{table:crossfoldsnv}
\end{table}

\begin{figure}[H]
\centering
\subfloat{\includegraphics[width=6.9cm]{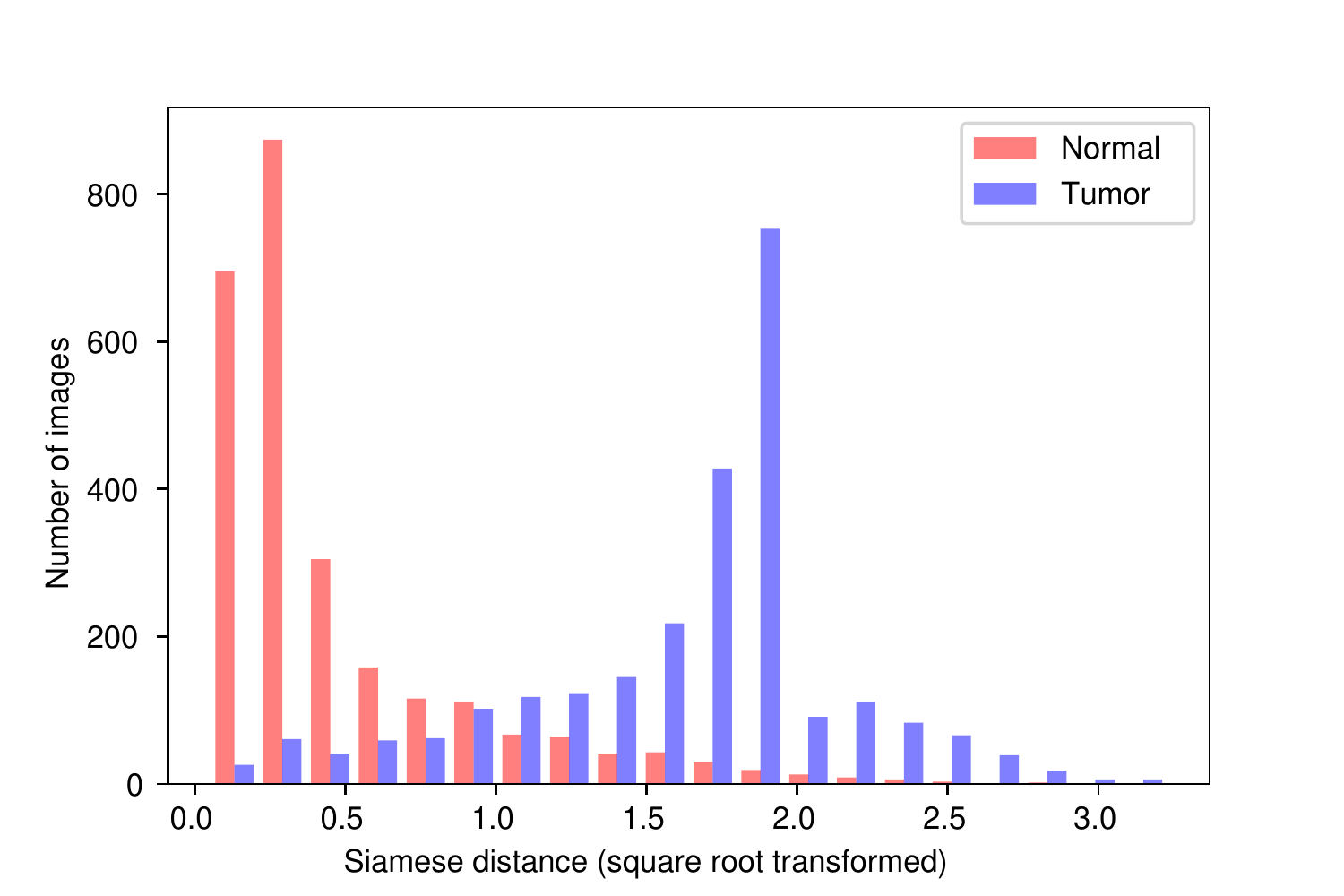}}
\subfloat{\includegraphics[width=6.9cm]{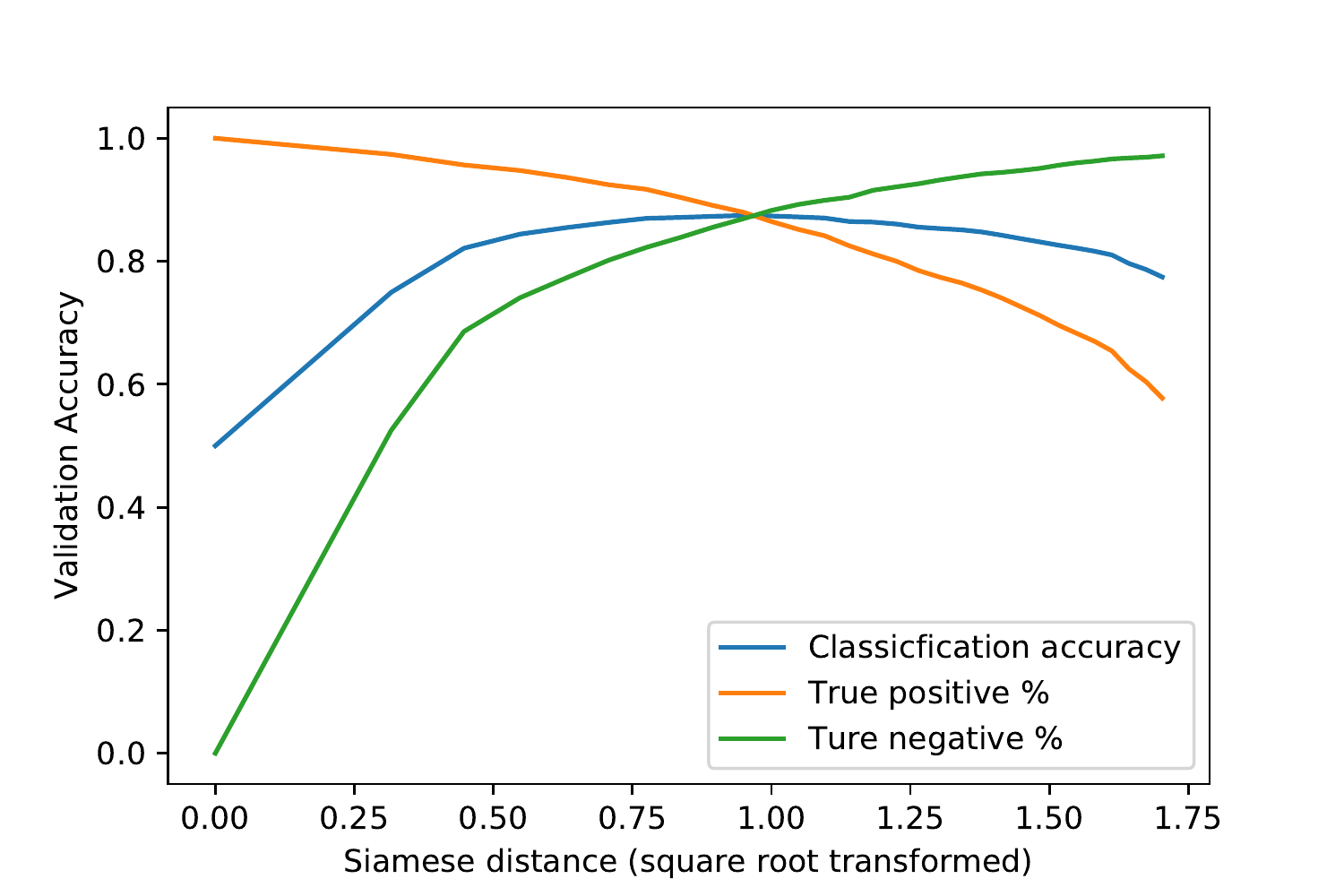}}
\caption{(1) Siamese distance distribution from tumor and normal tissue classes in the validation set, showing the model was able to learn to distinguish the two. (2) Validation accuracy vs different Siamese distances. The threshold of distances can be optimized to achieve high sensitivity and specificity when calling variants.}
\end{figure}

\section{Conclusion}

Our approaches provide promising novel machine-learning-based methods for improving genetic variant identification. These are used to both increase accuracy and to enable machine learning analysis of tumor/normal pairs in cancers. Additionally, these pile-up methods have the potential for use beyond just genomic variant detection, for example with alternative splicing events in transcriptomic pile-up images.

In conclusion, the methods presented herein describe a powerful framework for the application of machine learning technologies to genomic datasets. As an important first step in the development of a robust variant caller that outperforms traditional Bayesian approaches, we hope that these methodologies will eventually show clinical impact in the advancement of precision oncology.

\bibliography{bib}

\newpage
\appendix
\section{Appendix}
\subsection{A background of DNA sequencing, pile-ups, and encoding for CNNs}

When DNA is sequenced, the sequencer outputs data in the form of "reads". These are short stretches of DNA bases including Adenine (A), Cytosine (C), Guanine (G), and Thymine (T). These reads are then aligned to a human reference sequence to determine where in the genome they originated. The visualization of these aligned reads is known as a "pile-up", with the reference sequence along the top, and the reads stacked below. An individual's reads will differ slightly from the reference genome at DNA variant sites.

This pile-up can be encoded into an image, with each pixel as a DNA base. This in turn enables a CNN to classify the difference between true variants versus sequencing error (see schematic). This information can be encoded into a single channel (as below), or in RGB format to encode extra information into multiple channels (e.g. quality of the sequencing base call / likelihood of error of each base).

\begin{figure}[H]
 \label{fig:pileup}
 \includegraphics[width=1\linewidth]{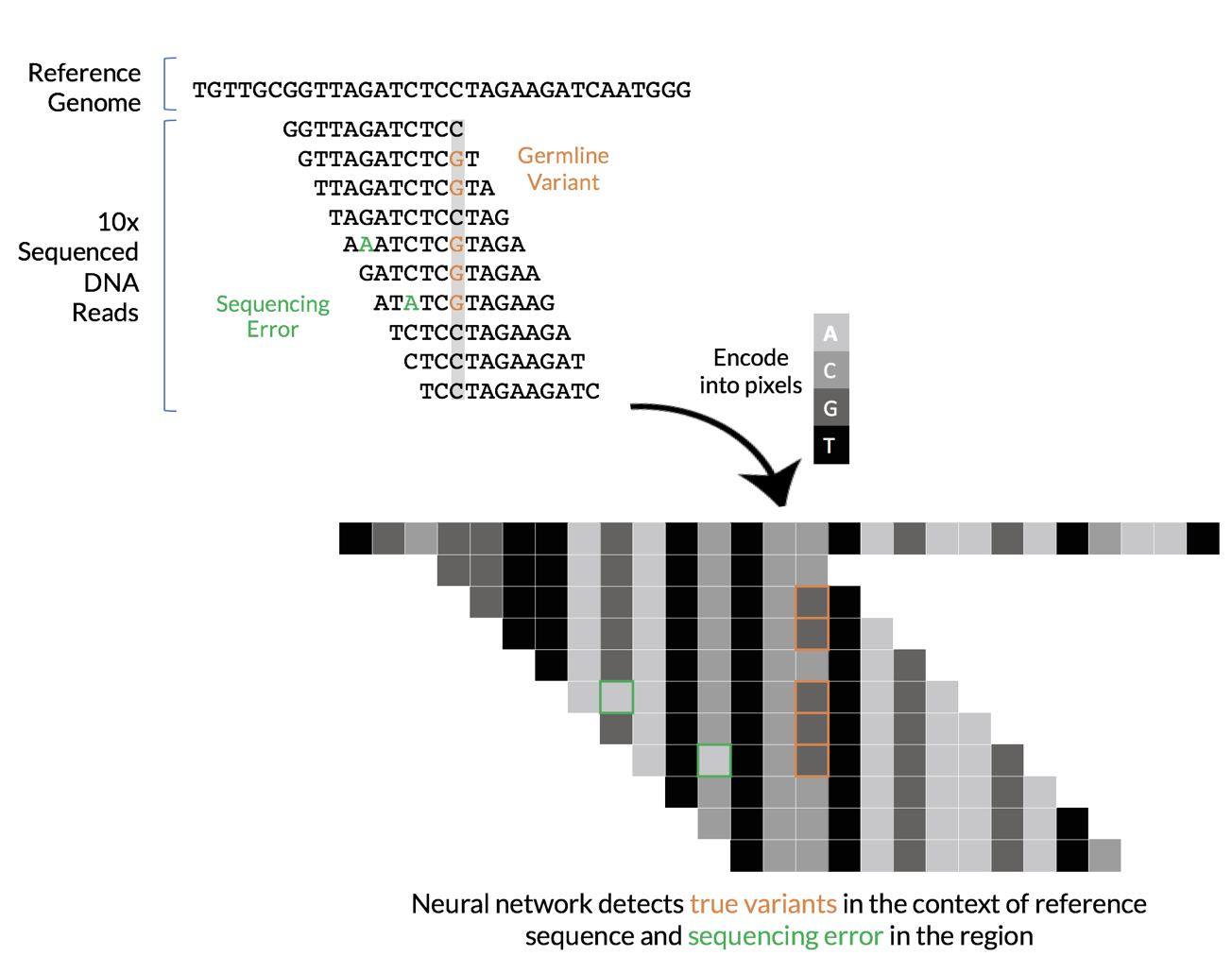}
\end{figure}

\end{document}